# EFFICIENT DISTANCE COMPUTATION ALGORITHM BETWEEN NEARLY INTERSECT OBJECTS USING DYNAMIC PIVOT POINT IN VIRTUAL ENVIRONMENT APPLICATION

Hamzah Asyrani Sulaiman[a], Abdullah Bade[b] & Mohd Harun Abdullah[c]

[a]*Universiti Teknikal Malaysia Melaka, 76100 Durian Tunggal, Melaka*
[b,c]*Universiti Malaysia Sabah, 88440 Kota Kinabalu, Sabah*
[a]*asyrani@utem.edu.my,* [b]*abade08@yahoo.com,* [c]*harun@ums.edu.my*

**Abstract.** Finding nearly accurate distance between two or more nearly intersecting three-dimensional (3D) objects is vital especially for collision determination such as in virtual surgeon simulation and real-time car crash simulation. Instead of performing broad phase collision detection, we need to check for accuracy of detection by running narrow phase collision detection. One of the important elements for narrow phase collision detection is to determine the precise distance between two or more nearly intersecting objects or polygons in order to prepare the area for potential colliding. Distance computation plays important roles in determine the exact point of contact between two or more nearly intersecting polygons where the preparation for collision detection is determined at the earlier stage. In this paper, we describes our current works of determining the distance between objects using dynamic pivot point that will be used as reference point to reduce the complexity searching for potential point of contacts. By using Axis-Aligned Bounding Box for each polygon, we calculate a dynamic pivot point that will become our reference point to determine the potential candidates for distance computation. The test our finding distance will be simplified by using our method instead of performing unneeded operations. Our method provides faster solution than the previous method where it helps to determine the point of contact efficiently and faster than the other method.

## INTRODUCTION

Colliding pairs in virtual environment is an event where two or more objects are trying to occupy the same space at the same time just before they penetrate with each other. In virtual environment application, such things are never meant to be existed since most of the application prevent that situation from happening by implementing collision detection system. Collision detection has become one of the most trendy research nowdays with various solution of implementation ranging from the usage of proximity queries and up to the hierarchy tree solution which is commonly known as Bounding-Volume Hierarchy (BVH).

Three-dimensional (3D) objects are uniquely exists as a virtual physical objects that had to enforce a collision detection system in order to efficiently exists with another 3D objects in virtual environment world. Collision detection system itself consisting three main components which are collision determination, collision detection and the collision response. Collision determination starts by pre determine the possible intersecting area by using approximation and then sending information to another phase of collision detection. Then, when the objects are really intersected between each other, collision detection will either performing broad phase checking that checks for early intersected area using Bounding-Volume (BV) and then proceed to use BVH for further testing. Then, when it comes to the leaf nodes of the BVH, which is the finest level of the intersected area, we need to perform narrow phase checking. Narrow phase checks for the object primitive itself which is the triangle by finding the possible triangle of one object with another object primitive. Then, we need to calculate the distance between both triangles and then check for the point of contact if both appears to having a contact. The last check is done by checking for the depth penetration that might occurs if the one of the object primitive penetrate the surface of the another object primitive. Collision response is the last process for simulating and animating what happens after the collision has been reported.

# BACKGROUND STUDIES

Performing collision detection between rigid bodies using primitive-primitive intersection checking required expensive computation cost. Some buildings might have thousands of polygons need to be checked for collision when some other objects undergoing intersection test with the buildings. By using brute force approach, each primitive will be tested for collision until the program found the correct intersection point. So, one of the way to handle this is to use a Bounding-Volume (BV). The purpose of using BV is to reduce the computational cost to detect object interference. If the object performs primitive-primitive testing without applying BV, it could consume longer time as it needs to check each triangle with other object triangle set. However, time to check for each collision can be reduced through enveloping highly complex object with BV. Instead of using single BV for one particular object in order to perform collision detection, BVH could help performing collision detection better than a single BV. BVH provides a hierarchical representation that could split the single BV into certain level before performing primitive-primitive testing for accurate collision detection. It also can be used for fast collision detection method by stopping at certain level using stopping function or criteria and approximately response to the collision as the object has been collided. It depends heavily on what kind of application that been developed as some application prefers speed and others accuracy.

At the present time, there are several famous BVs such as spheres [1], Axis Aligned Bounding Box (AABB) [2-4], Oriented Bounding Box (OBB) [4-6], Discrete Oriented Polytope (k-DOP) [7], Oriented Convex Polyhedra [8], and hybrid combination BV [9]. Most large scale 3D simulations used bounding box because of the simplicity, require less storage, fast response of collision, and easy to implement [10].

Distance Computation Algorithm for collision detection has been studied for the past three decades ago where M. Orlowski (1985) published a paper of "The Computation of the distance between polyhedra in 3-space", E.G. Gilbert, D.W. Johnson, and S.S. Keerthi (1988) published "A fast procedure for computing the distance between objects in three-dimensional space" and M.C. Lin (1991) published a popular paper of "A Fast Algorithm for Incremental Distance Calculation". Based on this paper, distance computation is mainly a method to determine the approximately high precision distance between pair of convex polyhedra. Several other implementation of using both techniques as a based technique are represented in [11-16].Distance computation algorithm is highly depends on the smallest step of object movement toward another objects as all computer simulation consists of coordination system. Thus, it has been used in another type of collision detection technique that focused on accuracy which is Continuous Collision Detection (CCD). Compared to Discrete Collision Detection (DCD), CCD provides a sequence of small, discrete steps that looks like a continuous movement.

# THEORITICAL FRAMEWORK

In order perform collision detection simulation between objects, we created an empty space environment with two objects which are Stanford Library 3D model. Figure 1 depicts our algorithm for computing distance between two near colliding triangle bound with AABB.

1. *Identify direction of object movement*
2. *Identify nearest midpoint of each object triangle that move toward the other.*
3. *Get minimum and maximum value of object movement axis*
    a. *Get first object minimum and maximum points*
    b. *Get second object minimum and maximum points*
4. *Construct Dynamic Inner AABB using 3(b) and 3(c)*
5. *Find midpoint and set as origin point*
6. *Calculate vertex to vertex and edge to vertex and edge to edge distance*
7. *Find shortest distance*

Figure 1: Dynamic Pivot Point Creation

The algorithm starts by identify the direction of object movement from two rigid bodies that each of triangle has successfully contain one triangle in one AABB. In this case, we will use an example of two-dimensional (2D) case where we need to compute the distance between these two near intersect triangles. Figure 2 shows object triangle bounded with an AABB.

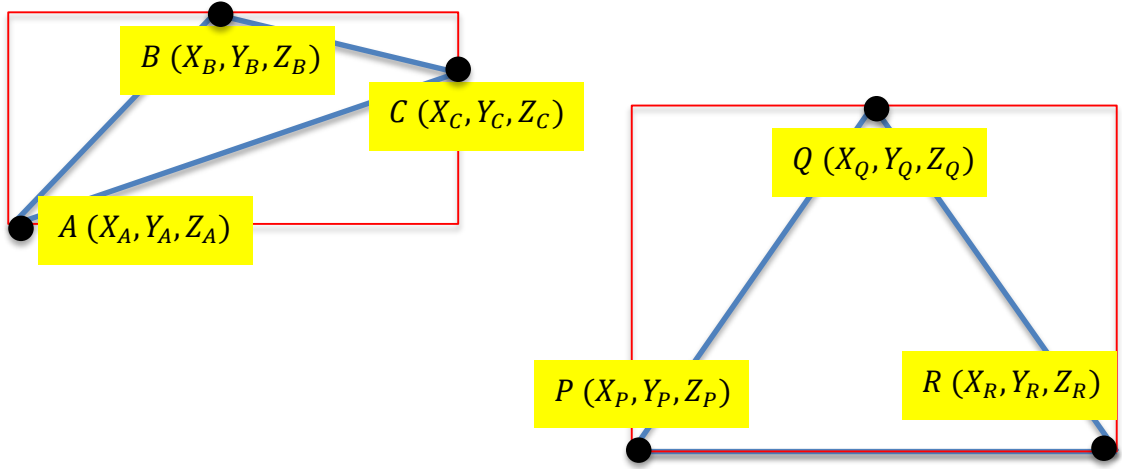

**Figure 2**: Performing distance computation between two triangles bound with AABB

Meanwhile, the final information to create an Internal AABB is to find out which triangle is on the higher location compared to the other one. In this case, we just need to find out whether $Y_A$, $Y_B$, $Y_C$, $Y_P$, $Y_Q$ and $Y_R$ have the minimum and the maximum point. For triangle ABC, $Y_B$ is the maximum Y coordinate while $Y_A$ is the minimum Y coordinate. Triangle PQR has $Y_Q$ as the maximum Y coordinate while both $Y_P$ and $Y_R$ can be choose as minimum Y coordinate. By using this information, we now can create another parallel lines which is in Y axis for both triangles. Since triangle ABC is in higher location compared to triangle PQR, thus we will use minimum Y coordinate point which is $Y_A$, and maximum Y coordinate of triangle PQR of $Y_Q$ to create our parallel lines. Internal AABB will be created based on this information of $X_C$ and $X_R$ and $Y_A$ and $Y_Q$. The same procedure apply if we have information on Z coordinate movement.

We now conclude that by obtaining three points from first triangle and another tree points from the second triangle we can reduce the testing instead of nine vertex-vertex testing into four vertex-vertex testing. Hence, by using an example of triangle ABC and PQR, we only need to check vertex-vertex testing between only two vertices for each triangle. Next, based on the figure 4, we illustrates dynamic origin point (DyOP) using Y and X of internal AABB.

Based on figure 3, we can obtain our midpoint of our internal AABB by using equations as shown in equation 1 and 2. This will become our dynamic origin point that keep on updating if both objects is moving.

$$X_{Mid\ Internal\ AABB} = \frac{X_P + X_C}{2} \qquad eq\ (1)$$

$$Y_{Mid\ Internal\ AABB} = \frac{Y_Q + Y_A}{2} \qquad eq\ (2)$$

# ANALYSIS AND DISCUSSION

In this experiments, we set up the environment by iterating a moving triangle with a static triangle. Each moving triangle will be tested against another shape of triangle. Given the name for each triangle is Obj1 until Obj10, thus we iterated Obj 1 with Obj 2, Obj1 with Obj3, Obj1 with Obj4, Obj1 with Obj5, Obj1 with Obj6, Obj1 with Obj7, Obj1 with Obj8, Obj1 with Obj9, and Obj1 with Obj10. A total of nine tests for Obj1. Where Obj1 became the moving triangle candidate while others become the static triangle for our experiments. For another iteration, we will repeat the process for Obj2, Obj3 until Obj10 and skipped any same Obj testing. In this case, no such thing as Obj1 with Obj1, Obj2 with Obj2, and others. For all triangles, a total 90 iterations/tests will be conducted.

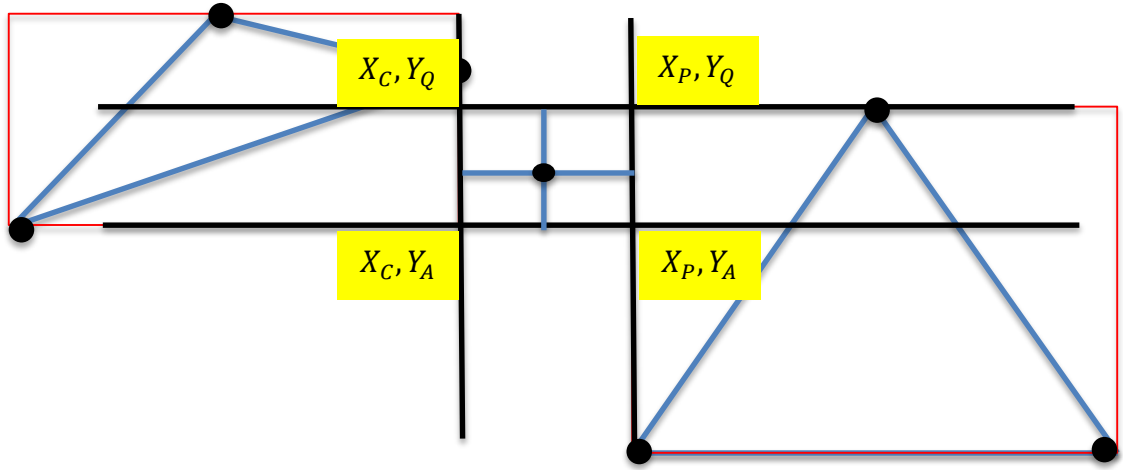

**Figure 3**: Internal AABB with dynamic origin point (DyOP) for calculating between point and edge or edge and edge by minimizing the efforts of primitive testing

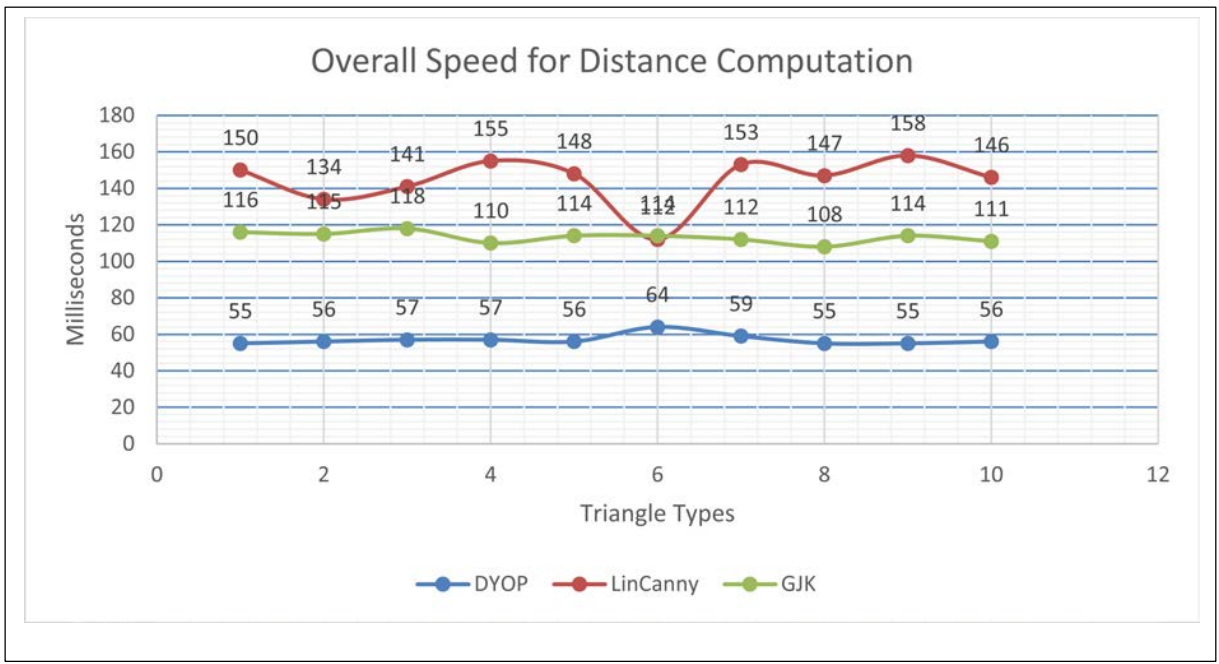

**Figure 4**: Overall Speed for Distance Computation

We also setup a fixed distance for all algorithm. Instead of randomization, we need to check the efficiency of the algorithm by maintaining only a single fixed distance. Thus, all the algorithm will be conducted their test using the same distance check. The time frame is captured for each nine iteration for each Obj1 until Obj10. Figure 4 shows our result of this experiments.

From figure 4, our proposed algorithm of DyOP performance is better than the other two algorithms in term of algorithm speed for distance computation checks. All algorithms used the same fixed distance and the same tests. By improving the speed of distance computation algorithm, we can potentially increase the speed of collision detection testing especially the continuous collision detection (CCD) types where distance computation algorithm is mainly used in CCD application such as medical simulation and high precision simulation. Which means, more testing will be conducted when the speed is improved and maintaining the accuracy of the collision detection algorithm. Figure 5.44 shows the differences in percentage for distance computation speed experiments.

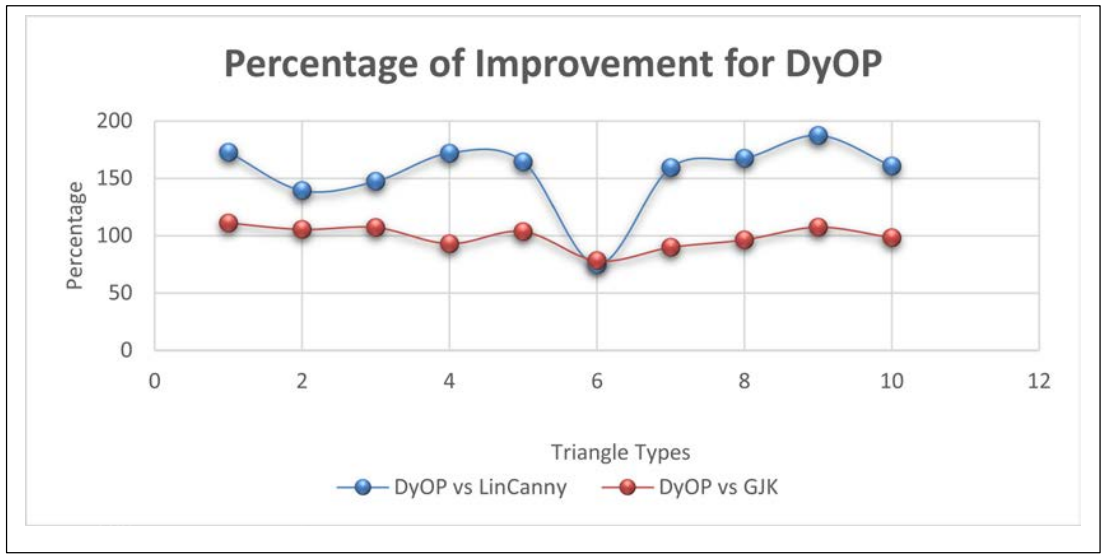

**Figure 5**: Overall Speed for Distance Computation

Based on figure 5, the highest of percentage between DyOP and LinCanny algorithm is 187.2727273% while the lowest being 75%. This means that our DyOP has proven to provide faster distance computation algorithm other than the other algorithms. We also have a 107.2727273% increase for DyOP versus GJK algorithm and the lowest being 78.125%. This proven enough to us that our algorithm works in better efficiency than the other algorithms.

## CONCLUSION AND FUTURE WORK

Based on all experiments, we concluded that our proposed algorithm of DyOP is more superior to the other two algorithms of GJK and Lin-Canny. The speed of computing distance is increase between ranges of 150% to 180% for certain type of triangles. Our DyOP algorithm is implemented in 2D environment where it is the easiest part to see the contribution of this algorithm. It is the fundamental idea that we need to chase down to the bottom of complex model which is a triangle in order to visualize properly what is the idea of DyOP compared to any other methods. Moreover, it is to give other researcher about an overview about this new and novel algorithm at the fundamental level before implement it into any other applications.

## ACKNOWLEDGMENTS

We would like to give special credit to Universiti Teknikal Malaysia Melaka for research grant under PJP/2013/FKEKK(15A)/S01199 for financial support. Also to the Universiti Malaysia Sabah for research alliance toward this research projects for supervision and monitoring.